# Thermoelectric Properties of Nanocomposite Heavy Fermion $CeCu_6$

Mani Pokharel[1]*, Tulashi Dahal[2], Zhifeng Ren[2] and Cyril Opeil[1]*

[1]*Department of Physics, Boston College, Chestnut Hill MA 02467, USA*

[2]*Department of Physics and TcSUH, University of Houston, Houston, TX 77204, USA*

*To whom correspondence should be addressed

**Abstract**

Heavy fermion compounds $CeCu_6$ were prepared by hot-press technique. Temperature-dependent (5 – 350 K) thermoelectric transport properties of the samples were measured. The dimensionless figure-of-merit (*ZT*) was optimized by varying the hot-pressing temperature. Our measurements of thermal conductivity show that the lowest hot pressing temperature (450 $^o$C) produces the lowest thermal conductivity. Electrical resistivity increases significantly while the Seebeck coefficient decreases with decrease in the hot pressing temperature. As the hot-pressing temperature decreases, electronic contribution to the total thermal conductivity decreased more rapidly than the lattice contribution did. As a result, for lower hot-pressing temperature the gain in thermal conductivity reduction was offset by the loss in power factor. Our ZT calculations show a broad peak with a maximum value of 0.024 at ~ 60 K for the sample hot pressed at 800 $^o$C. The pronounced low-temperature *ZT* peak emphasizes the importance of this heavy Fermion system as a potential *p*-type thermoelectric for solid state cooling applications.

**Introduction**

In recent years, solid-state cooling based on the Peltier effect has gained increased attention. A Peltier cooler with a high coefficient of performance (*Φ*) requires a material with a high value for the dimensionless figure of merit (*ZT*) defined by, $ZT = S^2\rho^{-1}\kappa^{-1}T$. Here *S* is the Seebeck coefficient, *ρ* is the electrical resistivity, *κ* is the thermal conductivity and T is the absolute temperature [1-3]. Optimizing the *ZT* of a material has been challenging due to the complex and interrelated quantities: *S*, *ρ* and *κ*. Current state of-the-art materials typically possess a *ZT* ≈ 1.5 at their peak operating temperature near or above 300 K. Such a system has yet to be discovered for low temperature applications. Yet, correlated electron systems are considered one of the classes of materials which might be useful as Peltier coolers below 77 K. Among the correlated systems, heavy-fermion compounds (HFCs) show promising thermoelectric properties at low temperatures, with a large *S* and small *ρ*. In these compounds, below some characteristic temperature $T_k$ a sharp peak of the density of states develops at the fermi level ($E_F$) which results in highly enhanced values for both the Sommerfeld (γ) and Seebeck coefficient (*S*).

Since the discovery by Stewart *et al.* [4] in 1984, $CeCu_6$ has been one of the widely studied HFCs. A great deal of interest was focused on transport properties of this system in the following years [5-10], owing to the Fermi liquid (FL) behavior at low temperature, similar to that of $CeAl_3$ [11]. Subsequent studies on this compound were focused on non-fermi liquid (NFL) behavior [12, 13] and anomalous thermopower [14-16]. In the recent years this compound has been a platform to study quantum critical point (QCP) behavior [17]. Although measurement of the thermoelectric properties of $CeCu_6$ had long been taken as an excellent approach to investigate quasiparticle excitation, little attention has been paid to its potential as a thermoelectric material for cooling purpose.

Nanostructuring has been proven to be very effective to reduce thermal conductivity without effectively degrading the electronic properties. Most studies focused on employing this technique desire to increase *ZT* near or above room temperature. Since the phonon contribution, in general, increases with a decrease in temperature; nanostructuring could be promising to reduce the thermal conductivity of thermoelectric material at low temperatures. In our previous work we successfully enhanced the *ZT* of strongly correlated narrow–gap semiconductor $FeSb_2$ using nanostructuring [18-20]. This paper represents the continuation of our studies concerned with improving *ZT* of a material using nanostructuring at low temperature (below 77 K). We present the result from our measurement of the thermoelectric properties of $CeCu_6$ samples which were prepared by ball milling and hot-pressing.

**Experimental**

Stoichiometric amounts of Ce (99.9%) and Cu (99.999 %) were mixed and arc-melted together on a water-cooled copper hearth in an argon atmosphere. To improve chemical homogeneity, the melted sample was flipped on the hearth plate and re-melted three times. The resulting ingot was etched in dilute nitric acid then ball milled under argon atmosphere for five hours to create nanopowder of $CeCu_6$. The nanopowder was then hot pressed for two minutes at 450, 600 and 800 °C applying a uniaxial pressure of 80 MPa. For simplicity, the samples are given names as listed in Table 1. Mass density of the pressed samples was determined using the Archimedes' method. X-ray diffraction (XRD) was performed on the fresh fracture of the samples. Scanning Electron Microscopy (SEM) was used to take the image of the samples. The as formed nanopower had an average initial diameter of ~ 200 nanometers. The Seebeck Coefficient (*S*), electrical resistivity ($\rho$), and thermal conductivity ($\kappa$) from 5 to 300 K were measured on a sample of typical dimensions of a 3x3x4 $mm^3$. A 2-point method in thermal transport option (TTO) of the Physical Property Measurement System (PPMS) was used to measure the thermoelectric properties. The horizontal rotator option of PPMS was used to measure Hall coefficient ($R_H$) of the samples with typical dimensions of 1×2×10 mm$^3$.

**Results and Discussion**

Figure 1 shows the XRD pattern for the samples. The peak positions confirm the orthorhombic crystal structure indicating that the ingot was alloyed in a single phase form. The crystal structure is retained in the hot pressed samples.

Figure 2 shows the SEM images for the samples. All the images were taken with the equal magnification of 1000 and under the identical parameter settings. Microstructures of the ingot and the sample HP 800 are not much different and no observable grains are found. Note the nearly equal values for the densities of the sample HP 800 and the ingot (Table 1). At the lower hot pressing temperature, however, a grain distribution is clearly seen with the grains tending to agglomerate. Porosity of the samples increases at the lower hot pressing temperatures.

Figure 3 shows the electrical resistivity of the samples as a function of temperature. All the samples exhibit similar resistivity profile typical of single crystal CeCu$_6$. Below 300 K, the resistivity decreases as the temperature decreases until it reaches a flat minimum. At lower temperatures a Kondo-like behavior emerges with a negative value for $\partial\rho/\partial T$. The resistivity then reaches a maximum at around 15 K before declining sharply with decreasing temperature, an indication of coherence development. Electrical resistivity of the 800 °C hot-pressed sample is slightly increased when compared to the parent ingot. A comparison among the hot-pressed samples shows that the electrical properties of CeCu$_6$ are greatly affected by varying HP temperature. With decreasing HP temperature the electrical resistivity increases significantly. When comparing HP 800 and 450 °C samples, we note at 60 K an increase in resistivity by a factor of 3.4. Such a drastic increase in the electrical resistivity might be attributed to the reduced grain size and the increased porosity.

Figure 4 shows the total thermal conductivity ($\kappa$) for the samples as a function of temperature. The thermal conductivity for polycrystalline samples of CeCu$_6$ was taken from Ref. [21]. The thermal conductivity follows temperature dependence similar to that reported for another HFC CeCu$_4$Al in Ref. [22]. The total thermal conductivity decreases as the HP temperature decreases. At 50 K, κ was reduced from ~ 5 Wm$^{-1}$K$^{-1}$ (for ingot) to ~ 2 Wm$^{-1}$K$^{-1}$ (for sample HP 450), a reduction by 60 %. In general, $\kappa = \kappa_l + \kappa_e$, where $\kappa_l$ and $\kappa_e$ are the lattice and electronic contributions to the total thermal conductivity respectively. Generally, phonon-grain boundary scattering mechanism reduces the phonon contribution ($\kappa_l$) whereas porosity has shown to reduce the electronic contribution ($\kappa_e$) [23, 24]. Looking at the SEM images (Fig. 2), the reduction of the thermal conductivity with decrease in HP temperature might be attributed to the combined effect of both the contributions from phonon-grain boundary scattering and porosity effect.

In Figure 5 we present the temperature dependence of the Seebeck coefficient. All the samples exhibit a positive value for the Seebeck coefficient below 300 K with a maximum at $T_{max} \approx 50$ K. This value for $T_{max}$ is in agreement with the previously reported data [15, 16]. In the context of heavy-fermions, such a peak in $S$ at higher $T$ ($T > T_K$) is usually attributed to the Kondo scattering on higher multiplets (as opposed ground state doublet) which are split by crystal field effects (CEF). For $T > T_{max}$, $S$ follows an unusual temperature dependence of the form: $S \propto -lnT$. Whereas for T < $T_{max}$, $S$ follows T-behavior typical of metals. The Seebeck coefficient decreases as the HP temperature decreases.

We also measured the Hall coefficient ($R_H$) of the samples at 60 K, temperature at which the $ZT$ curve peaks. Under the assumption of single parabolic band (SPB) model, the effective carrier density ($n$) and the Hall mobility ($\mu$) were calculated using the formulas, $n = 1/|R_H|e$ and $\mu_H = |R_H|/\rho$ respectively, where e = 1.6 ×$10^{-19}$ C is the magnitude of electronic charge. $R_H$, $n$ and $\mu_H$ of the samples are listed in Table 1. Magnitude of the Hall coefficient for the ingot sample (7.12 × $10^{-4}$ $cm^3C^{-1}$) is of the same order as reported in literature [25]. The Hall coefficient and the Hall mobility decrease by two orders of magnitude going from the ingot to the hot-pressed samples. $R_H$ and $\mu_H$ decrease with the hot-pressing temperature which is consistent with the resistivity and Seebeck coefficient data.

Figure 6 shows the temperature dependence of the dimensionless figure-of-merit ($ZT$). The $ZT$ values assume a peak at around 60 K for all the samples. The peak value of $ZT$ for the optimized sample HP 800 is 0.024 at 60K. Since the ingot and the sample HP 800 have comparable values of the power factor at 60 K (Inset of Fig 6), the improved $ZT$ basically comes from the thermal conductivity reduction. Here we note that he value for $ZT$ greater than 0.1 at cryogenic temperatures (< 77 K) has rarely been reported. $FeSb_2$ single crystal exhibits a peak ZT value of ~ 0.005 at ~ 10 K [26] which was increased to 0.013 at 50 K in nanostructured samples [18]. Single crystal FeSi has $ZT$ of 0.01 at 50 K, which can be slightly raised to 0.07 at 100 K by 5% Ir doping [27]. $CeB_6$ is one of the best thermoelectric materials at low temperature for which ZT = 0.25 at 7 K was reported [28].

While the original goal for nanostructuring was to increase the power factor (*PF*) employing quantum confinement of carriers [29, 30], experiments [31-33] have shown that the key reason for improved $ZT$ was the reduction of thermal conductivity. Therefore in the recent years researches on nanostructured thermoelectric material are focused on reducing the thermal conductivity, while producing minimal adverse effects on the related parameters of Seebeck coefficient and electrical conductivity. This approach seems to work effectively mainly for the system in which the thermal transport is phonon-dominated (as opposed to electron-dominated). One of the ways to analyze the effectiveness of nanostructuring is to look at the values of the reduced Lorenz number ($L/L_0$). Here $L$ is defined as $L = \kappa\rho T^{-1}$ and $L_0 = 2.45 \times 10^{-8}$ $W\Omega K^{-2}$ is the free-electron value. In general value of $L/L_0$, much greater than 1 implies that the phonons are dominant. In Figure 7, we have presented the temperature dependence of $L/L_0$. The shape of the $L/L_0$ $(T)$ curve for all the samples is typical of heavy fermions. $L/L_0$ decreases in the sample HP 800 when compared to the ingot meaning that the lattice contribution to the total thermal conductivity was effectively reduced in the sample HP 800. However with a further decrease in the HP temperature, $L/L_0$ increases significantly. At lower HP temperatures, the thermoelectric properties are affected in such a way that the electronic contribution to the total thermal conductivity decreases more rapidly than the phononic contribution does. As a result no net gain in *ZT* was achieved by lowering the hot-pressing temperature.

Our improvement in *ZT* for the 800 $^o$C hot-pressed sample, as compared to the ingot, results from a combination of nanostructuring and optimization of its hot-press/annealing temperature. This lowers the thermal conductivity maintaining a relatively good electrical conductivity therefore enhancing *ZT* at low temperatures. Although a *ZT* of 0.024 at ~ 60 K may appear low when compared to other thermoelectric *ZT*'s at the same temperature, e.g. Bi88Sb12, it is nonetheless significant because of its positive Seebeck coefficient. The broad and pronounced peak in *ZT*, shown in Fig. 7, emphasizes its potential as a good *p*-type candidate for low-temperature solid state cooling applications. Further enhancement of *ZT* is possible by doping on either the Ce or Cu sites which may decrease resistivity further and promote phonon scattering.

**Conclusion**

In conclusion, nanostructured samples of $CeCu_6$ were prepared by ball milling of arc-melted ingot followed by hot pressing. The thermoelectric properties were optimized varying the hot pressing temperature. The thermal conductivity decreased as the hot pressing temperature decreased showing that nanostructuring is an effective approach to reduce thermal conductivity of this system. However, the electrical properties were degraded adversely to decrease the power factor. Overall the *ZT* values decreased with decrease in hot pressing temperature. A significant value for the *ZT* of 0.024 at 60 K was observed for the optimized sample HP 800. The broad and pronounced peak in *ZT* emphasizes its potential as a good *p*-type thermoelectric material at low-temperature. Further improvement of *ZT* of this system by combining nanostructuring and doping can be expected.

**Acknowledgment**

We gratefully acknowledge funding for this work by the Department of Defense, United States Air Force Office of Scientific Research's MURI program under contract FA9550-10-1-0533.

**References**

1. H. J. Goldsmid, *Thermoelectric Refrigeration* (Plenum Press, New York, 1964)

2. D. M. Rowe. Ed. *CRC Handbook of Thermoelectrics* (CRC Press. Boca Raton, FL, 1995)

3. T. M. Tritt, Ed. *Semiconductors and Semimetals* (Academic Press, San Diego, CA 2001)

4. G. R. Stewart, Z. Fisk, M. S. Wire, *Phys. Rev*. B **30** (1984)482

5. Y. Onuki, Y. Shimizu, T. Komatsubara, *J. Phys. Soc. Japan* **53** (1984) 1210

6. A. Amato, D. Jaccard, E. Walker, J. Flouquet, *Sol. State Commun*. **55** (1985) 1131

7. Y. Onuki, Y. Shimizu, T. Komatsubara, *J. Phys. Soc. Japan* **54** (1985) 304

8. Y. Onuki and T. Komatsubara, *J. Magn.Magn. Mater*., **63** and **64** (1987) 281

9. F. P. Milliken, T. Penney, F. Holtzberg, Z. Fisk, *J. Magn. Magn. Mater*. **76** (1988) 201

10. A. Amato, D. Jaccard, J. Flouquet, F. Lapierre, J. L. Tholence, R. A. Fisher, S. E. Lacy, J. A. Olsen, N. E. Phillips, *J. Low Temp. Phys*. **68** (1987) 371

11. H. R. Ott, H. Rudigier, Z. Fisk, J.O. Willis, and G. R. Stewart, *Solid State Commun*. **53** (1985) 235

12. H. V. Lohneysen, T. Portisch, H. G. Schlager, A. Schroder, M. Sieck, and T. Trappmann, *Phys. Rev. Lett*. **72**, 20 (1994)

13. A. Rosch, A. Schroder, O. Stockert, and H. v. Lohneysen, *Phys. Rev. Lett*. **79**, 1 (1997)

14. J. Sakurai, F. Taniguchi, K. Nishimura, K. Sumiyama, H. Amano, and K. Suzuki, *Physica* B 186-188 (1993)

15. M. Ocko, M. Miljak, I Kost, J-G Park, and S. B. Roy, *J. Phys. Condens. Matter* **7** (1995) 2979-2986

16. M. I. Ignatov, A. V. Bogach, G. S. Burkhanov, V. V. Glushkov, S. V. Demishev, A. V. Kuznetsov, O. D. Chistyakov, N. Yu. Shitsevalova, and N. E. Sluchanko, *Journal of Experimental and Theoretical Physics*, **105** (2007) 58-61

17. N. E. Sluchanko, D. N. Sluchanko, N. A. Samarin, V.V. Gluskhov, S. V. Demishev, A. V. Kuznestov, G. S. Burkhanov, and O. D. Chistyakov, *Low Temperature Physics* **35**, 7 (2009).

18. H. Zhao, M. Pokharel, G. Zhu, S. Chen, K. Lukas, Q. Jie, C. Opeil, G. Chen, and Z. Ren, *Appl. Phys. Lett*. **99**, 163101 (2011)

19. M. Pokharel, H. Zhao, R. Lukas Z., and C. and Opeil, *Mater. Res. Soc. Symp. Proc. Vol. **1*** © 2012

20. H. Zhao, M. Pokharel, S. Chen, B. Liao, K. Lukas, C. Opeil, G. Chen, and Z. Ren, *Nanotechnology* **23**, 505402 (2012).

21. Y. Peysson, B. Salce, and C. Ayche, *J. magn. Magn. Mater* **54-57** (1986) 423-424

22. M. Falkowski, and A. Kowalczyk, *Proceedings of the European Conference Physics of Magnetism* (PM'11), Pozman, June 27-July 1, 2011

23. H. Lee, D. Vashaee, D. Z. Wang, M. S. Dresselhaus, Z. F. Ren and G. Chen, *J. Appl.Phys*. **107**, 094308 (2010)

24. J.M. Montes, F.G. Cuevas and J. Cintas, *Appl. Phys*. A **92**, 375–380 (2008)

25. A. S. Krivoshchekov, B. N. Goshchitskii, V. I. Voroin, I. F. Berger, Yu. N. Akshentsev, and A. E. Earkin, *Physica* B **359-361** (2005) 178-180

26. A. Bentien, S. Johnsen, G. K. H. Madsen, B. B. Iversen, F. Steglich, *Europhys. Lett.*, **80** (2007) 17008

27. B. C. Sales, E. C. Jones, B. C. Chakoumakos, J. A. Fernandez-Baca, H. E. Harmon, J. W. Sharp, E. H. Volckmann, *Phys. Rev. B: Condens. Matter.*, **50** (1994) (12), 8207-8213.

28. S. R. Harutyunyan, V. H. Vardanyan, A. S. Kuzanyan, V. R. Nikoghosyan, S. Kunii, K. S. Wood, and A. M. Gulian, *Appl*. *Phys*. *Lett*. 83 (2003) 2142

29. L. D. Hicks and M. S. Dresselhaus, *Phys*. *Rev*. B, **47** (1993), pp. 12727

30. L. D. Hicks and M. S. Dresselhaus, *Phys*. *Rev*. B, **47** (1993), pp. 16631

31. R. Venkatasubramanian, E. Silvola, T. Colpitts, and B. O'Quinn, *Nature*, **413** (2001), pp. 597-602.

32. T. C. Harman, P. J. Taylor, M. P. Walsh, and B. E. LaForge, *Science*, **297** (2002), pp. 2229-2232.

33. K. F. Hsu, S. Loo, F. Guo, W. Chen, J. S. Dyck, C. Uher, T. Hogan, E. K. Polychroniadis, and M. G. Kanatzidis, *Science*, **303** (2004), pp. 818-821.

**Table 1:** Hall coefficient ($R_H$), carrier density (*n*) and Hall mobility (μ) for the samples at 60 K.

| Sample name | Hot-pressing temperature (°C) | Hall coefficient (cm³/C) | Carrier density (cm⁻³) | Hall mobility (cm²V⁻¹S⁻¹) | Mass density (gcm⁻³) |
|---|---|---|---|---|---|
| Ingot |  | $7.12 \times 10^{-4}$ | $8.78 \times 10^{21}$ | 14.12 | 7.48 |
| HP 800 | 800 | $5.18 \times 10^{-5}$ | $1.20 \times 10^{23}$ | 0.85 | 7.55 |
| HP 600 | 600 | $4.47 \times 10^{-5}$ | $1.39 \times 10^{23}$ | 0.43 | 6.62 |
| HP 450 | 450 | $2.61 \times 10^{-5}$ | $2.38 \times 10^{23}$ | 0.13 | 6.51 |

**Figure Captions**

**Figure 1:** X-ray diffraction pattern for the arc melted ingot and the three hot pressed samples of CeCu$_6$.

**Figure 2**: SEM images of the samples.

**Figure 3:** Electrical resistivity as a function of temperature for the CeCu$_6$ samples.

**Figure 4:** Thermal conductivity as a function of temperature for the three nanostructured CeCu$_6$ samples. The data for the polycrystalline sample was drawn from Ref. [21] and replotted for comparison.

**Figure 5:** Seebeck coefficient as a function of temperature for the CeCu$_6$ samples.

**Figure 6:** *ZT* as a function of temperature for the CeCu$_6$ samples. Inset shows the power factor as a function of temperature.

**Figure 7:** Reduce Lorenz number as a function of temperature for the CeCu$_6$ samples. *L* is defined as $L = \kappa\rho T^{-1}$ and $L_0 = 2.45 \times 10^{-8}$ WΩK$^{-2}$ for free electron was used in calculation.

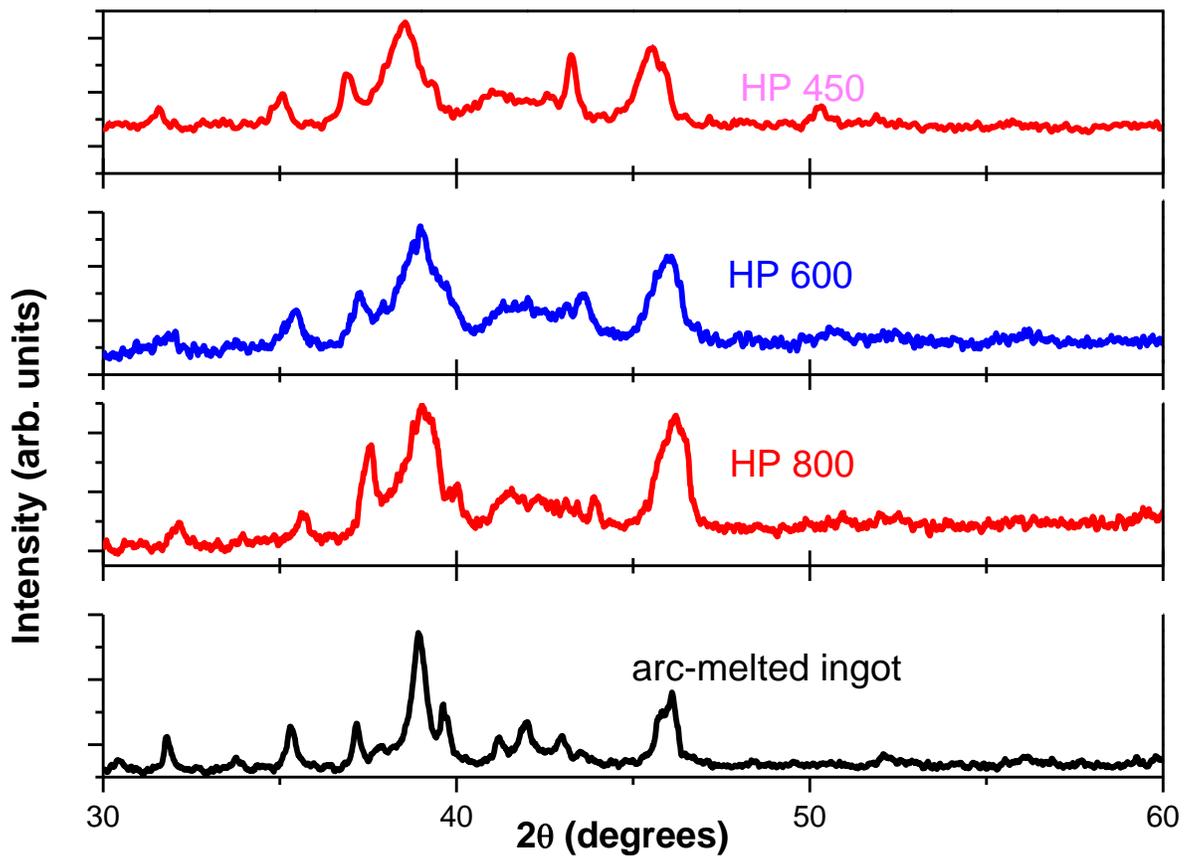

Figure 1

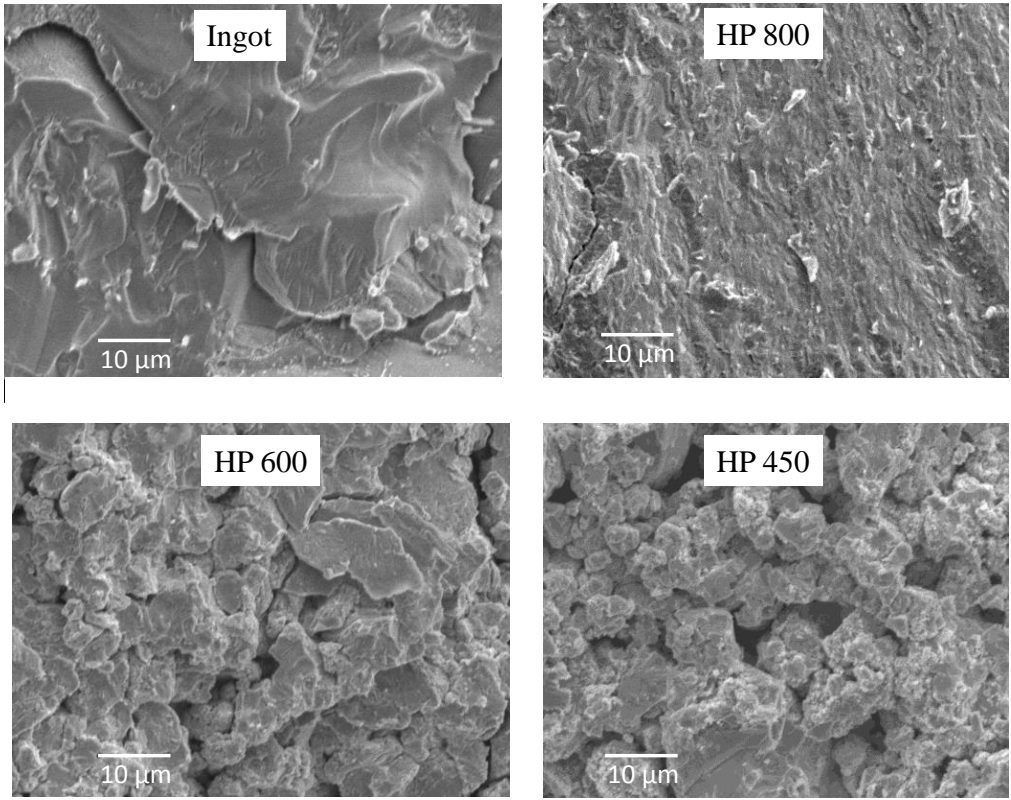

Figure 2

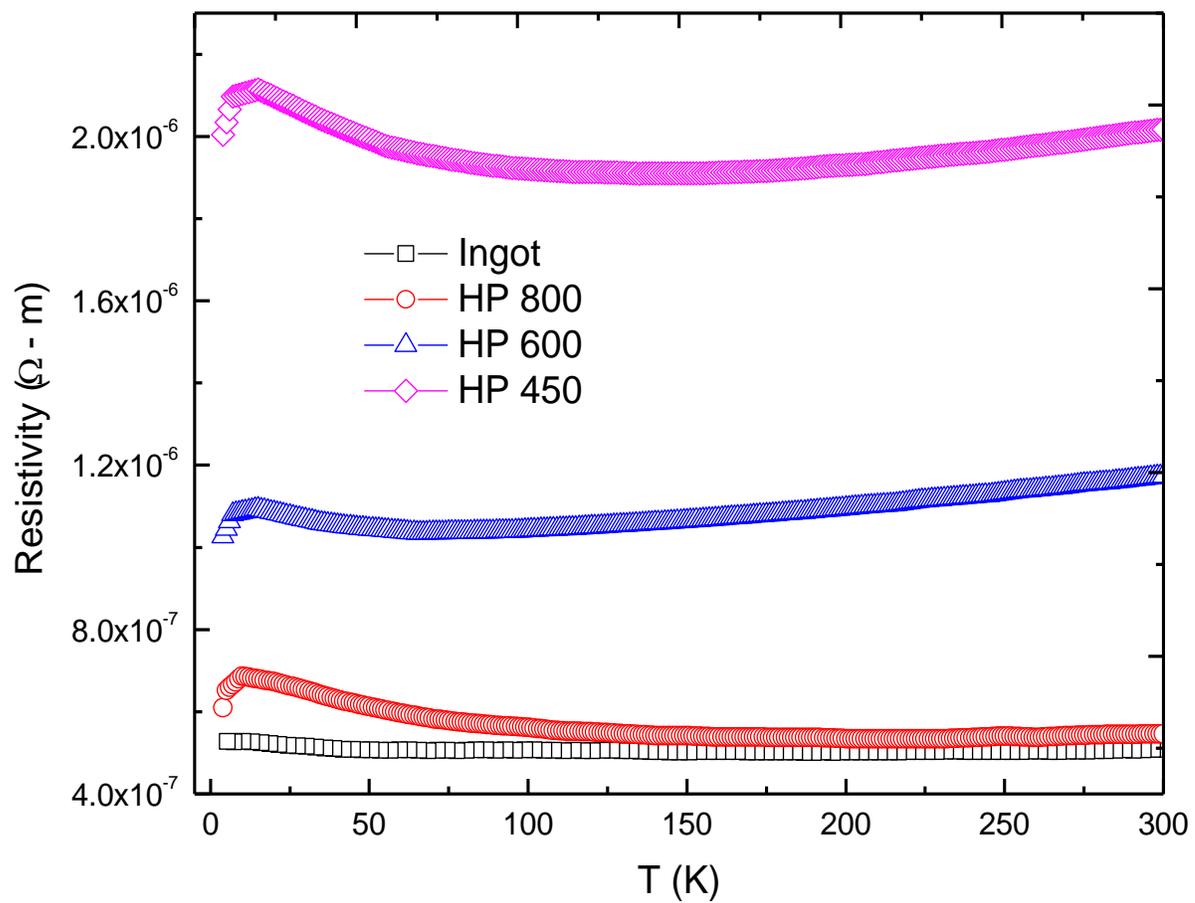

Figure 3

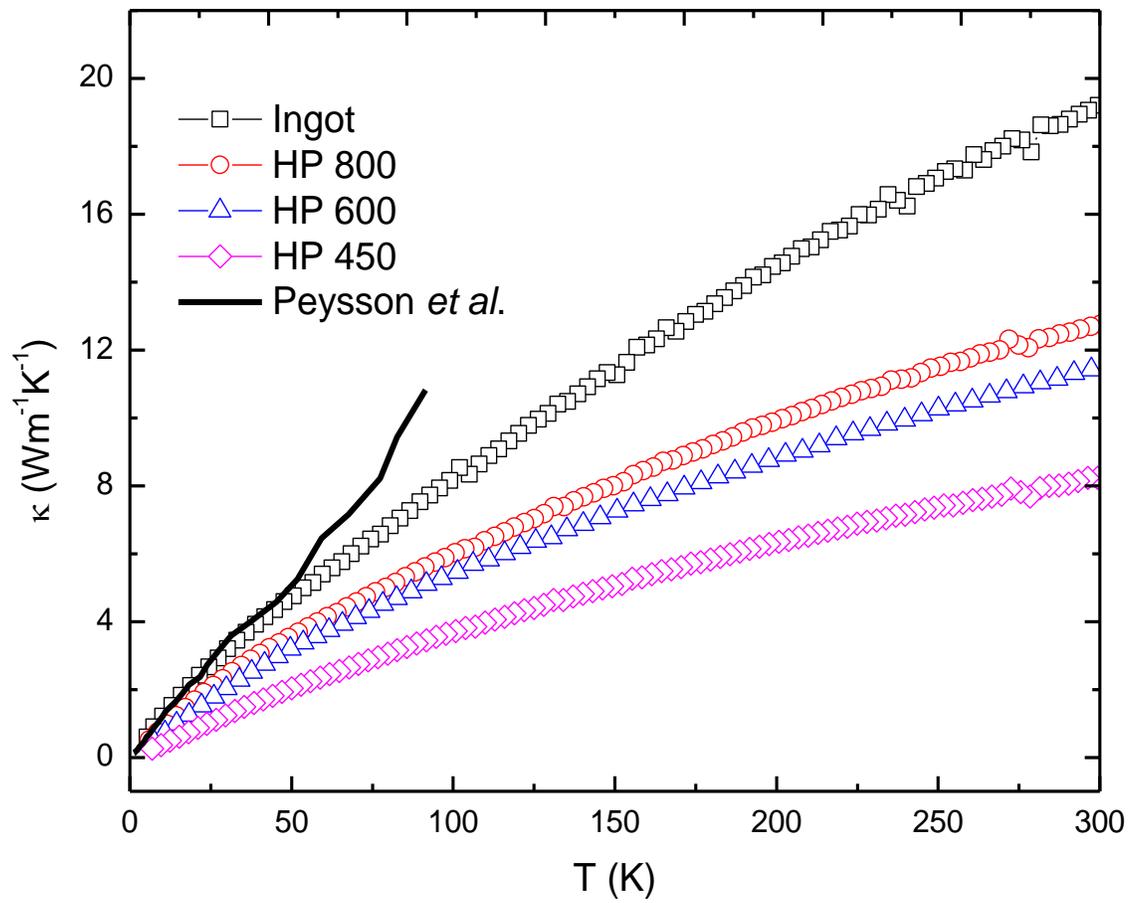

Figure 4

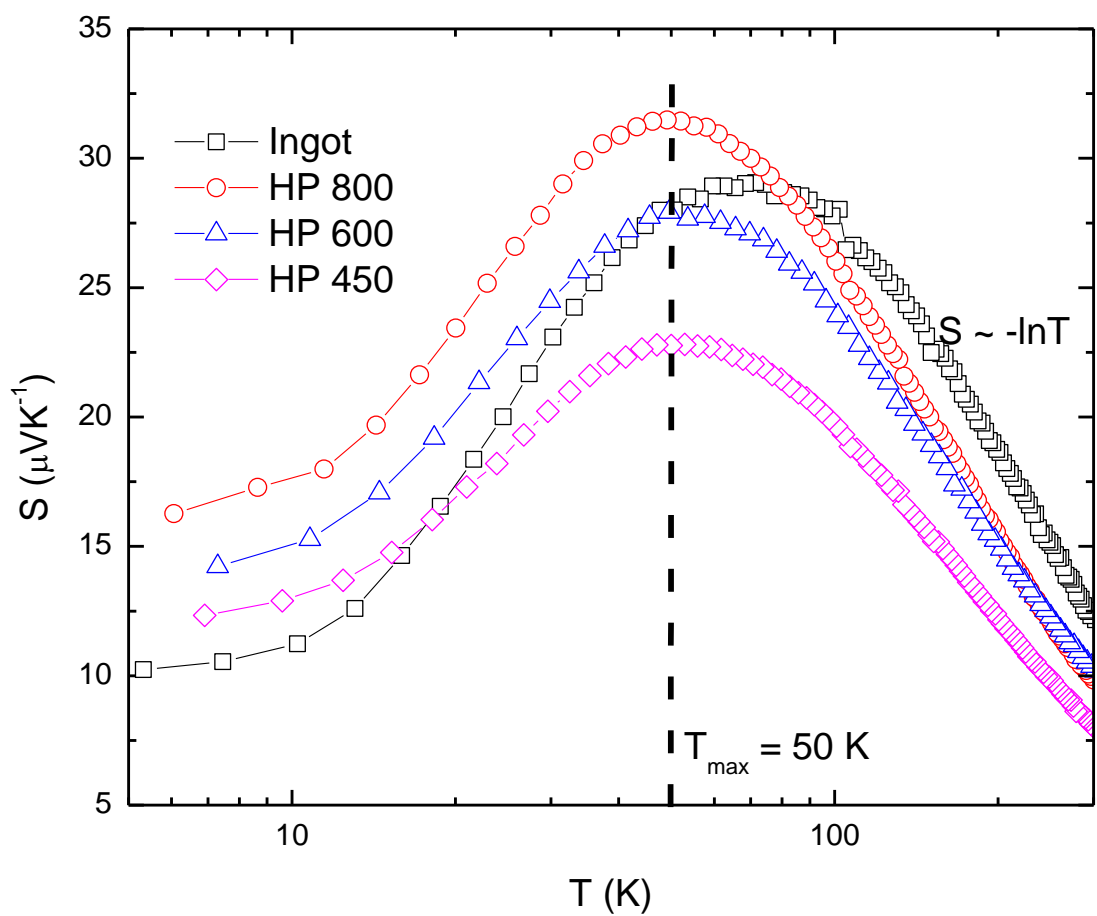

Figure 5

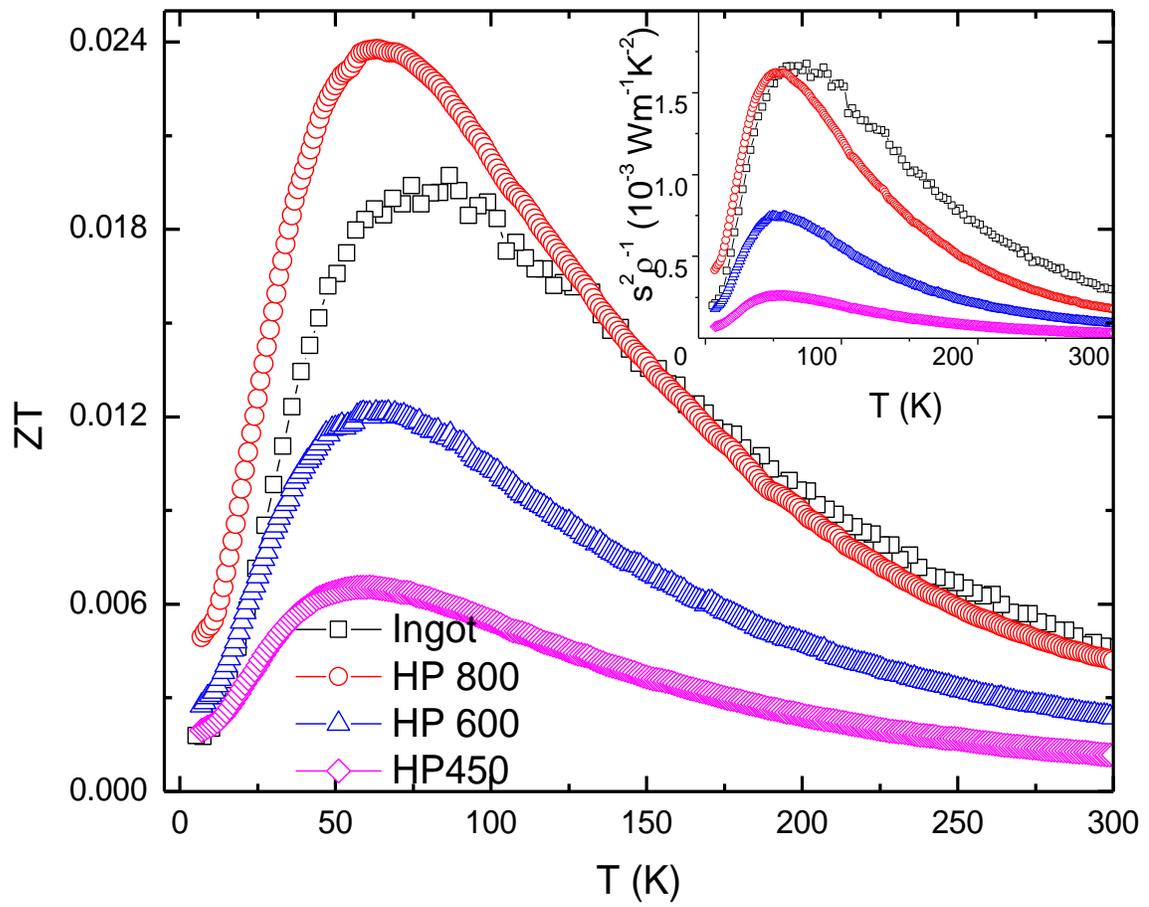

Figure 6

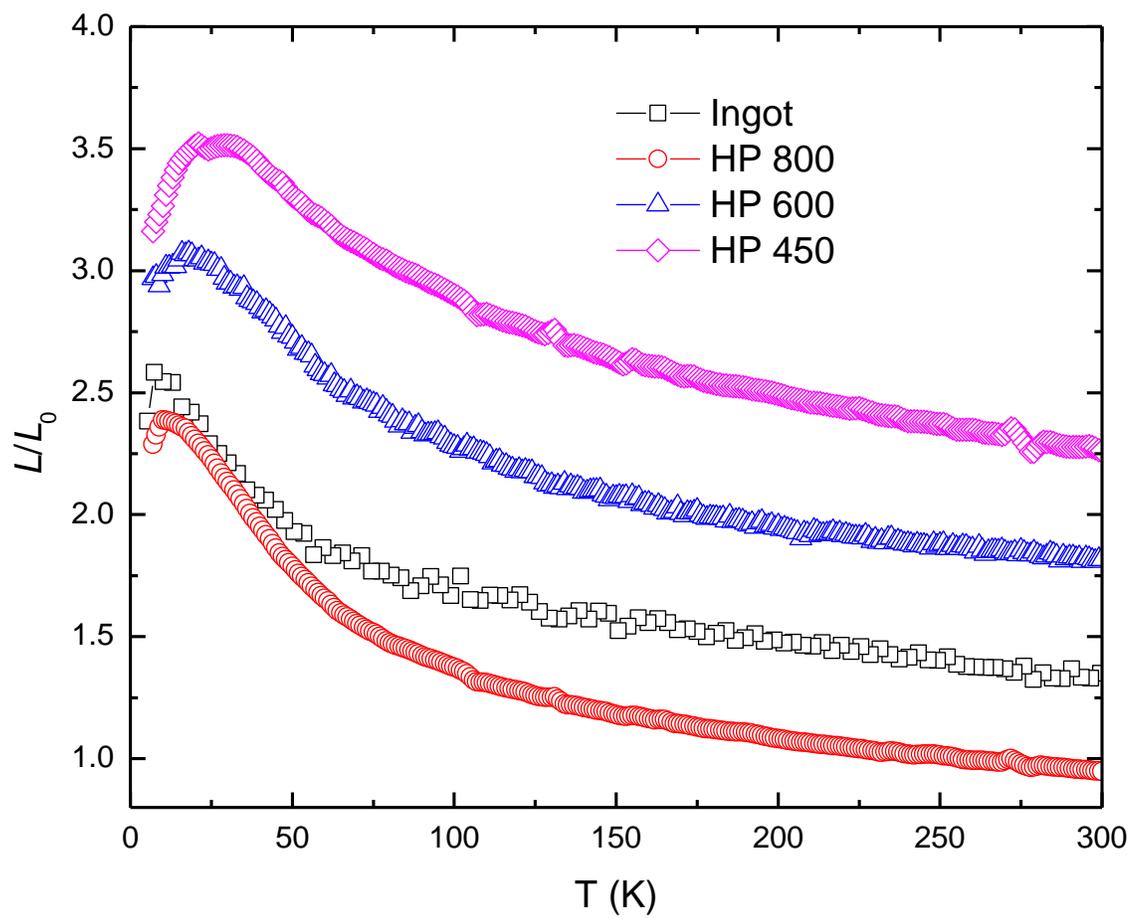

Figure 7